\documentclass[showpacs,aps]{revtex4}
\usepackage{amsmath}
\usepackage{amsfonts}
\usepackage{amssymb}
\usepackage{graphicx}
\usepackage{subfigure} 
\usepackage{bm} 

\begin{document}
\title{Short Pulse Dynamics in Strongly Nonlinear Dissipative Granular Chains}
\author{Alexandre Rosas}
\affiliation{
Departamento de F\'{\i}sica,
Universidade Federal da Para\'{\i}ba,
Jo\~ao Pessoa, Para\'{\i}ba, Brazil, Caixa Postal 5008 - CEP: 58.059-970}
\author{Aldo H. Romero} 
\affiliation{
Cinvestav-Quer\'etaro,
Libramiento Norponiente 200, 
76230, Fracc. Real de Juriquilla,
Quer\'etaro, Quer\'etaro, M\'exico}
\author{Vitali F. Nesterenko}
\affiliation{
Department of Mechanical and Aerospace Engineering,
University of California San Diego,
La Jolla, CA 92093-0411}
\author{Katja Lindenberg}
\affiliation{
Department of Chemistry and Biochemistry,
and Institute for Nonlinear Science
University of California San Diego
La Jolla, CA 92093-0340}
\date{\today}

\begin{abstract}
We study the energy decay properties of a pulse propagating in a strongly nonlinear 
granular chain with damping proportional to the relative velocity of the grains.
We observe a wave disturbance that at low viscosities consists of two parts exhibiting two entirely
different time scales of dissipation.  One part is an attenuating solitary wave, 
is dominated by discreteness and nonlinearity effects as in a dissipationless chain,
and has the shorter lifetime.  The other is a purely dissipative shocklike structure with a
much longer lifetime and exists only in the presence of dissipation.
The range of viscosities and initial configurations that lead
to this complex wave disturbance are explored.
\end{abstract}


\maketitle

\section{Introduction}
Granular chains under impulse loading are known to support a rich variety of
excitations~\cite{nesterenko,nesterenkobook,hinch,senreview,ourpre,ourprl}.  The
precise nature of the excitations depends on a number of features that include
the particular granular configurations, state of precompression, dimensionality, shape of
initial disturbance and, of particular interest to us in this work,
viscosity.  Consider first the behavior of a monodisperse one-dimensional chain
in which the granules are
placed side by side, just touching but without precompression. A velocity $v_0$
imparted to a single
grain quickly evolves into a solitary wave carrying energy whose dynamical
evolution depends on the
nature of the granules.  In particular, for elastic spherical grains one obtains
a
solitary wave that resides on about 5-7 granules and whose velocity depends on
its amplitude~\cite{nesterenko}.

We are interested in the effect of viscosity on the propagation of such an
initial excitation.
There are a number of different sources and descriptions of viscous
effects~\cite{duvall,brunhuber,arevalo,brilliantov,sen,falcon,ourpretwogran}, and
in~\cite{ourpre} we studied the dynamics of a pulse in one such case, when the
granular chain is
immersed in a viscous medium that gives rise to a Stokes drag proportional to grain velocity. 
Here we consider the more prelavent situation in which the viscosity
arises from the interaction between the grains as one grain rubs against
another, expanding on our earlier results for this case~\cite{ourprl,herbold1,herbold2}.
Here the dissipative contributions are proportional to the relative velocities of grains in elastic
contact. It should be noted that in experiments on a chain immersed in some liquids the main
dissipative contribution was also found to be proportional to the relative velocities of grains
(rather than a Stokes drag term) due
to the expulsion of liquid from the area of developing elastic contact~\cite{herbold1}.
While the former viscous interaction leads to energy \emph{and} momentum loss to
the medium, the
latter involves only internal dissipative forces and is consequently momentum conserving. 
If there were only binary collisions in the chain, the momentum conserving dissipation would be one possible
dynamical description of the usual parametrized coefficient of restitution~\cite{landau}.

Our model is the simplest example of a strongly nonlinear discrete system that
can be verified
experimentally.  It consists of a chain of granules that interact via the purely
repulsive power law
potential
\begin{equation}
\begin{array}{l l l}
V(\delta_{k,k+1})&=\frac{a}{n}|\delta|^{n}_{k,k+1}, \qquad &\delta\leq 0,\\ \\
V(\delta_{k,k+1})&= 0, \qquad &\delta >0,
\end{array}
\label{eq:hertz}
\end{equation}
where 
\begin{equation}
\delta_{k,k+1} \equiv y_k - y_{k+1}.
\end{equation}
Here $a$ is a prefactor determined by Young's modulus $E$, the Poisson ratio
$\sigma$, and the
principal radius of curvature $R$ of the surfaces at the point of
contact~\cite{landau,hertz}; $y_k$ is the displacement of granule $k$ from its
equilibrium position.
The exponent $n$ depends on the shapes of the contacting surfaces. For spherical
granules
(Hertz potential) $n=5/2$ and $a=[E/3(1-\sigma^2)]\sqrt{2R}$~\cite{hertz}.
The force between two grains is nonzero only when the grains are in contact, and
consists of the 
mechanical force which is the negative derivative of the potential, and a
viscous force that is
proportional to the relative velocity of the interacting granules.  We
introduce the rescaled position $x_k$, time $t$, and viscosity coefficient
$\gamma$ used in Refs.~\cite{ourpre,ourprl},
\begin{equation}
x_k=\frac{y_k}{b}, \qquad t =\frac{v_0\tau}{b},\qquad \gamma = \tilde{\gamma}
\frac{b}{mv_0};\qquad
b\equiv \left( \frac{mv_0^2}{a}\right)^{1/n},
\end{equation}
where $y_k$, $\tau$, and $\tilde{\gamma}$ are the corresponding unscaled
quantities. 
The equation of motion for the $k$th grain then takes the form
\begin{equation}
\ddot{x}_k = \left[\gamma \left(\dot{x}_{k+1} -\dot{x}_k\right) - (x_k -
x_{k+1})^{n-1}\right] \theta (x_k -
x_{k+1}) +  \left[\gamma \left(\dot{x}_{k-1} -\dot{x}_k\right) + (x_{k-1} -
x_{k})^{n-1}\right] \theta (x_{k-1} -
x_{k}),
\label{eq:motion}
\end{equation}
where a dot denotes a derivative with respect to $t$.  The Heaviside function
$\theta(y)$ ensures
that the elastic and the viscous grain interactions exist only if the grains are in
contact.
Note that in this scaled equation of motion the constant $a$ as well as the mass
have been scaled
out, and the initial velocity imparted to a grain is now unity.
In the absence of viscosity an initial impulse given to one grain of
the chain of otherwise resting grains placed side by side quickly settles into a
forward propagating stationary pulse that is increasingly
narrow with increasing $n >2$. As we will show, and as one might expect, the
viscosity results in an overall
exponential energy dissipation. In a viscous medium, if a Stokes drag is the dominant source
of dissipation, this is essentially all that the viscosity
does~\cite{ourpre}, so that the pulse simply keeps moving without changing its
shape but with a
decreasing amplitude (and consequently velocity).  Remarkably,
in the present situation with dissipation determined by the relative velocity of grains
the behavior
is entirely different. Below a critical viscosity, there is a steady removal of energy 
from the pulse, part of which contributes to the formation of a long tail created
behind it from which energy is dissipated much more slowly.  The removal of
energy from the leading
pulse is rapid because, being very narrow, it has large velocity gradients, The
long tail that
evolves behind it loses energy far more slowly because the velocity gradients
there are very
small~\cite{ourprl}.

Exact analytic solutions to the equations of motion~(\ref{eq:motion}) appear
unattainable, and so in this work we rely heavily on numerical simulations, with analytic arguments
where possible.  In Sec.~\ref{sec2} we present a detailed study of the propagation of an impulse
along a viscous chain when the granules are initially placed side by side with no gaps and no
precompression. In Sec.~\ref{sec3} we explore the effects of an initial precompression
on the propagation of the impulse, and briefly comment on the effects of initial gaps.  A summary of
our findings is presented in Sec.~\ref{sec4}.

\section{Dynamics with no precompression and no gaps}
\label{sec2}
\begin{figure}
\begin{center}
\includegraphics[width=10cm,height=6.0cm]{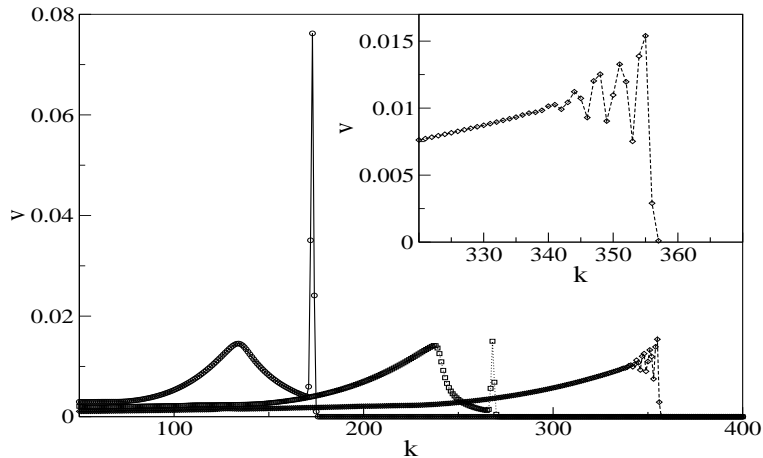}
\caption{
Time snapshots of the velocity profile for small viscosity ($\gamma=0.02$).
The progression of the profile is easily recognizable as the composite pulse
moves forward, the secondary pulse steepens and the primary pulse
disappears.  The times are $500$, $900$, and $1400$ and $n=5/2$.
Inset: detailed view of the crest of the velocity profile at time
$1400$.
\label{fig:snap_g002}}
\end{center}
\end{figure}

Consider a granular chain in which the granules are placed side by side,
just touching but without
precompression.  This situation is referred to in Ref.~\cite{nesterenkobook} as a ``sonic vacuum"
because such a system does not support sound waves.  In the next section we will relax this condition. A 
velocity is imparted to a single grain at one end of the chain. 
Below a critical viscosity, after a short time in which a small amount of energy is lost
through some back-scattering of nearby granules in the wake, almost all of the impact energy
resides in a forward traveling wave that has an unusual two-part structure which is formed very
rapidly and comes about as follows~\cite{ourprl}.  A pulse similar to a narrow solitary wave caused
by the strongly nonlinear forces in the discrete medium is generated.  This ``primary pulse," being
spatially very narrow, exhibits high velocity gradients that lead to a loss of
its energy. Some of the lost energy is simply dissipated, but part of the energy of the leading
pulse goes into a
quasistatic compression that appears behind the primary pulse because the forward displacements
of the particles closer to the impacted end are larger than the corresponding displacements of
particles further away from this end due to the attenutation of the velocity of the particles in the
propagating wave. This compression, and the associated pulse that arises
as a result (``secondary pulse"), are entirely dependent on the presence of dissipation. As a
result, this two-wave structure is not observed in the usual dissipationless models.
The secondary pulse is very broad and is therefore far more persistent than the primary pulse
because it has much smaller velocity gradients. 
A typical progression with time of the velocity profile for a small viscosity is shown in
Fig.~\ref{fig:snap_g002}. 
The figure exhibits all the characteristics 
discussed above, specifically, the narrow decaying primary pulse and the evolving secondary pulse.
A more detailed discussion of this figure will be presented later.

\begin{figure}
\begin{center}
\includegraphics[width=10cm,height=6cm]{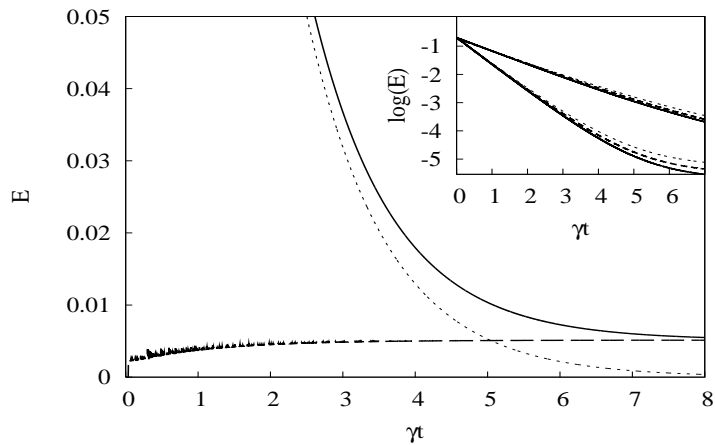}
\caption{Breakdown of the total energy (solid line) into portions
associated with the primary pulse (dotted) and the secondary pulse
(dashed) for $n=5/2$ and $\gamma=0.01$. Inset: time evolution of the
total energy, where two sets of three curves are shown, corresponding
to $n=2.2$ (upper set) and $2.5$ (lower set) in the Hertz potential.
Each set shows results for $\gamma=0.01$ (dotted), $0.005$ (dashed),
and $0.001$ (solid).
The early rapid decay mainly reflects
the rapid energy loss of the primary pulse.  The remaining energy
stored in the secondary pulse dissipates on a much longer time scale. 
\label{fig:breakdown}}
\end{center}
\end{figure}

The total energy of the system as a function
of time is shown in the inset in Fig.~\ref{fig:breakdown} for
two values of $n$ and three values of $\gamma$.
In our scaled units, the initial energy is $1/2$.
The attenuation of the energy of the ``primary" and ``secondary"
portions for the case $n=5/2$ and $\gamma=0.01$ is shown in
Fig.~\ref{fig:breakdown}, demonstrating the separation of time scales
for energy dissipation.  The primary pulse is a highly
non-stationary portion of the wave that
maximizes the rate of dissipation of some of the energy, as
reflected in the steep exponential decay associated with this loss. 
The energy decay slows down drastically as the primary pulse vanishes and
only the more persistent secondary pulse remains. Dissipation of
energy in the rapid decay regime is faster for
higher $n$. This behavior is due to the larger velocity
gradients in the primary pulse whose width decreases with increasing
$n$~\cite{nesterenkobook}.

To set the stage for further discussion, we recall the successful treatment of this problem in the
absence of dissipation based on a long wavelength
approximation~\cite{nesterenko,nesterenkobook,hinch,ourpre}.  This analysis,
based on a continuum description, is extremely successful 
despite the fact that the pulse is very
narrow in the underlying discrete system, i.e., in spite of the fact that a continuum approximation 
assumes that the particle characteristic length is much smaller
than the wavelength, which in fact it is not.
For large $n$, when the pulse is extremely narrow, a
decidedly discrete binary collision approximation works even better~\cite{ourprepulsevel}, but
the continuum approximation
still works well at least qualitatively. Some of the effects of discreteness are
captured by expanding $x_{k\pm 1}$ about $x_k$ in a Taylor series up to fourth order (the usual
second order or diffusive expansion would of course not capture any such effects).  The continuum
equation is obtained by neglecting the one-sidedness of the potential, that is, ignoring the fact that
there are no
attractive forces in the problem.  This works because the solution is assumed to apply only to the
compressed portion of the chain, and to be identically equal to zero outside of this region.  The
equation of motion for $n>2$ obtained in this way is~\cite{nesterenko,nesterenkobook}
\begin{equation}
\frac{\partial^2 x}{\partial t^2}
= \frac{\partial}{\partial k} \left [- \left ( - \frac{\partial x}
{\partial k}\right )^{n-1} + \frac{n-1}{24}
\left ( - \frac{\partial x}{\partial k}\right )^{n-2}
\frac{\partial^3 x}{\partial k^3} \right ] - \frac{1}{24}
\frac{\partial^3}{\partial k^3} \left [\left (
- \frac{\partial x}{\partial k}\right )^{n-1} \right ].
\label{eq:nestwavegen0}
\end{equation}
This equation admits a solution of the form~\cite{nesterenko},
\begin{equation}
\left (-\frac{\partial x}{\partial \xi}\right ) =
A_0 \sin^m \alpha \xi,
\label{eq:soliton}
\end{equation}
where $\xi = k - c_0 t$ and $c_0$ is the pulse velocity. Direct substitution of
this solution into Eq.~(\ref{eq:nestwavegen0}) leads to the values
\begin{equation}
m=\frac{2}{(n-2)}, \qquad
\alpha=\left[\frac{6(n-2)^2}{n(n-1)}\right]^{1/2},\qquad c_0=\left(\frac{2}{n}\right)^{1/2}
A_0^{\frac{n-2}{2}}.
\label{eq:constants}
\end{equation}
A solitary wave is constructed by retaining this solution over one period, $0\leq \alpha(k-c_0t)\leq
\pi$, and setting $\partial x/\partial \xi$ equal to zero outside of this range. This solution does
not satisfy the velocity pulse initial condition because it is meant to describe the system after a
short initial transient whereupon it settles into this traveling configuration.  One can go further
and take advantage of the fact that almost all of the initial energy resides in this pulse (an
extremely small portion is lost to back scattering~\cite{hinch,ourpre}).  Using conservation of energy
arguments and dealing carefully with the fact that due to nonlinearity the kinetic ($K$)  and potential
($U$) energies are not equal but instead obey a generalized equipartition theorem~\cite{tolman},
one further finds that $K/U=n/2$ so that $K=n/[2(n+2)]$. The kinetic energy can be calculated
directly by explicit integration, 
\begin{equation}
K = \int_0^{\pi/2\alpha} \dot{x}(\xi) \mathrm d \xi = \frac{c_0^2
A_0^2}{2\alpha} I\left(\frac{4}{n-2}\right),
\end{equation}
where
\begin{equation}
I(s) = \int_0^{\pi} \sin^s \theta \mathrm d \theta = 2^s \frac{\Gamma^2
\left(\frac{s+1}{2}\right)}{\Gamma\left(s+1\right)}.
\end{equation}
This result together with the generalized equipartition theorem then leads to explicit values for
$c_0$ and $A_0$.  In particular, for spherical granules we find $c_0=0.836$ and
$A_0=0.765$~\cite{ourpre}.

In the presence of dissipation, a similar expansion of the equations of motion~(\ref{eq:motion})
leads to an additional contribution in the continuum problem,
\begin{equation}
\frac{\partial^2 x}{\partial t^2} - \gamma \frac{\partial^2}{\partial
k^2}\left(\frac{\partial x}{\partial t}\right)
= \frac{\partial}{\partial k} \left [- \left ( - \frac{\partial x}
{\partial k}\right )^{n-1} + \frac{n-1}{24}
\left ( - \frac{\partial x}{\partial k}\right )^{n-2}
\frac{\partial^3 x}{\partial k^3} \right ] - \frac{1}{24}
\frac{\partial^3}{\partial k^3} \left [\left (
- \frac{\partial x}{\partial k}\right )^{n-1} \right ].
\label{eq:nestwavegen}
\end{equation}
Unfortunately, we have not found an exact solution to this equation, especially one that
at low viscosities captures
the two contributions to the solution whose features we have described on the basis of numerical
simulations.  In the following two subsections we discuss the dynamics of this solution mainly on
the basis of simulations.  At low viscosities, this discussion is reasonably organized into
a separate discussion for each of the two portions of the excitation, the primary and the secondary pulses,
because their evolution and decay involves such disparate time scales.  However, we wish to stress
that in spite of this separation, the pulse is one entity that
consists of two interdependent parts rather than a superposition of two independent entities.
Furthermore, clearly identifiable primary and secondary pulses occur only if the viscosity is sufficiently
small.  This point, as well as the behavior at higher viscosities, will be described 
in more detail below.

\subsection{Primary pulse}

\begin{figure}
\begin{center}
\includegraphics[width=10cm]{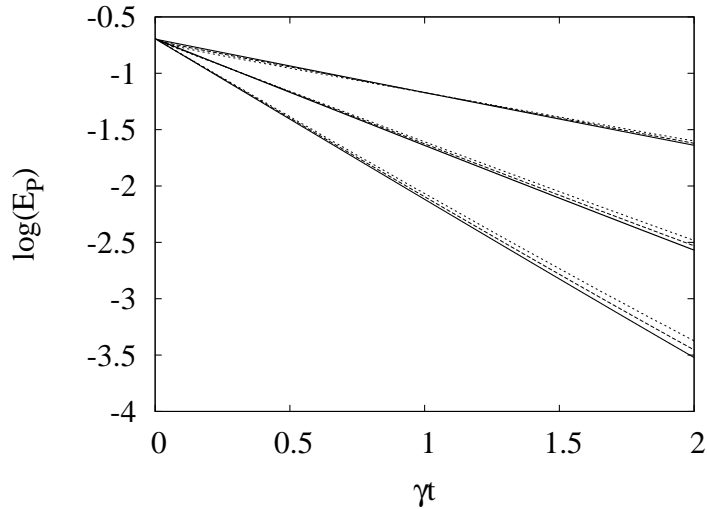}
\caption{
The pulse energy decays exponentially, as $e^{-2u\gamma t}$, with time. From top
to bottom, the three groups of lines correspond to $n=3, \; 5/2$ and $2.2.$ For
each $n,$ there are three values of $\gamma$: $0.01$ (dotted), $0.005$ (dashed)
and $0.001$ (filled). The slopes of the curves change slightly with $\gamma$
but are equal within a 10\% margin ($u=0.20\pm0.02$ for $n=2.2, \; u =
0.46 \pm 0.02,$ for $n=5/2$ and $u=0.70 \pm 0.03$ for $n=3$).
\label{fig:pulseenergy}}
\end{center}
\end{figure}

As in the dissipationless case, the primary pulse forms quickly following the initial velocity
impact due to the strong nonlinearity and dispersion.  The pulse travels along the chain with
diminishing speed (and with an evolving tail) since its amplitude decreases due to dissipation.
As noted above and as is apparent in
Fig.~\ref{fig:breakdown}, the loss of energy due to dissipation is initially essentially
exponential as long as the energy is mainly stored in the primary pulse.  While we know that it can
not be the entire solution because it does not account for the secondary pulse, we assume a solution
of the form Eq.~(\ref{eq:soliton}) but with a time dependent amplitude and velocity.  
This captures the behavior of the primary pulse if it retains its shape as it loses energy, and
is at best expected to hold as long as there is a prominent primary pulse, that is, for
times $\gamma t \lesssim 1$:
\begin{equation}
\left (-\frac{\partial x}{\partial \xi}\right ) =
A(t) \sin^{\frac{2}{n-2}} \alpha \xi(k,t),
\label{eq:with}
\end{equation} 
where
\begin{equation}
\xi(k,t) = k - \int_0^t c(t) \mathrm{d} t
\label{eq:xiprime}
\end{equation}
and
\begin{equation}
c(t) = \sqrt{\frac{2}{n}} A^{\frac{n-2}{2}}(t). 
\end{equation}
We may speculate that this approach is valid as long as the strong nonlinearity and dispersion are
more or less balanced as in a nondissipative chain, which is the case for the primary pulse.
To choose a specific functional form for $A(t)$ 
we must rely on the simulation results and observe that to times of order $\gamma t \sim 1$
the decay of the energy, as exhibited in Fig.~\ref{fig:breakdown}, is exponential.  This is confirmed
in more detail in Fig.~\ref{fig:pulseenergy}, where it is evident that for all the values of $n$ and
$\gamma$ in the figure the decay is well described by an exponential.  Furthermore, the slope is
fairly insensitive to the value of $\gamma$, so at least to a first approximation the energy decays as
$e^{-2 u\gamma t}$ where $u$ is a constant (and the $2$ is introduced for convenience).  
While the notation $E$ in Fig.~\ref{fig:breakdown} indicates the total energy and the notation 
$E_P$ in Fig.~\ref{fig:pulseenergy} indicates the primary pulse energy, at the early times of
Fig.~\ref{fig:pulseenergy} they are essentially the same.
The decay of the amplitude and the pulse velocity with the assumed form Eq.~(\ref{eq:with})
then are
\begin{equation}
A(t)= A_0 e^{-\frac{2u}{n}\gamma t}, \qquad c(t)=c_0
e^{-\frac{(n-2)u}{n}\gamma t},
\label{eq:A}
\end{equation}
and therefore
\begin{equation}
\xi(k,t) = k - c_0 \frac{n}{u\gamma (n-2)} \left (1 -
e^{-\frac{(n-2)u}{n}\gamma t} \right ).
\label{eq:xinest}
\end{equation}

\begin{figure}
\begin{center}
\includegraphics[width=10cm]{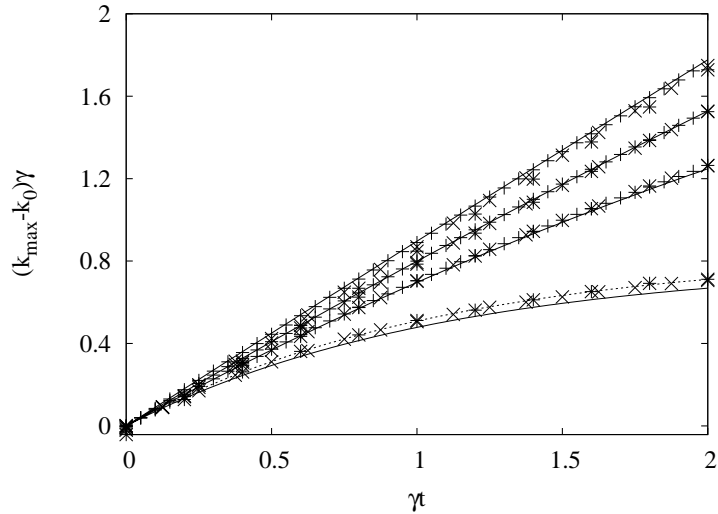}
\caption{Position of the grain with maximum velocity as a function of time for
the same values of $n$ (upper three groups) and $\gamma$ (0.001 plus signs, 0.005 crosses and 0.01 stars)
as in Fig.~\ref{fig:pulseenergy}. The lowest group of curves is for $n=7$.
Here, $k_0 = \frac{\pi}{2\alpha}.$ The solid lines are the theoretical predictions according to
Eq.~\ref{eq:kmax}.  The dotted curve for $n=7$ is obtained from the binary collision
approximation.
\label{fig:kmax}}
\end{center}
\end{figure}

The validity of the assumed form for the primary pulse can be tested numerically in a number of
ways.  For instance, our trial solution predicts that
the position of the maximum of the pulse should be~\cite{ourpre}
\begin{equation}
k_{max}=\frac{\pi}{2\alpha} +\frac{c_0}{\gamma u} \frac{n}{(n-2)} \left(
1-e^{-\frac{n}{(n-2)}u\gamma t}\right).
\label{eq:kmax}
\end{equation}
In Fig.~\ref{fig:kmax} we compare this prediction with the
numerical results. According to Eq.~(\ref{eq:kmax}), the curves
corresponding to different values of $\gamma$ should collapse for each $n$, and
indeed we have a clear collapse. Moreover, using the values of $u$ obtained from
the energy decay, we observe excellent agreement. 
As a test of the continuum approximation concept, we have included results for the very high value
$n=7$.  Here we observe a deviation between the numerical results (symbols) and the solid curve
obtained from Eq.~(\ref{eq:kmax}) with $u$ determined from the energy decay curve and $c_0$ obtained
via the continuum approximation.  For this large value of $n$ the pulse is extremely narrow, and so
we tested an alternative binary collision approximation in which we assume that only two granules
are involved in a collision at any one time~\cite{ourprepulsevel}. With $c_0$ calculated from this
approximation we obtain the dotted line, which is in excellent agreement with the simulations.  We
stress that in any case the curve collapse predicted by the form~(\ref{eq:kmax}) is observed for all
the values of $n$ and $\gamma$ exhibited in the figure.  This confirms that the pulse position is
indeed an exponential function of $\gamma t$.

\begin{figure}[ht]
\begin{center}
\includegraphics[width=10cm]{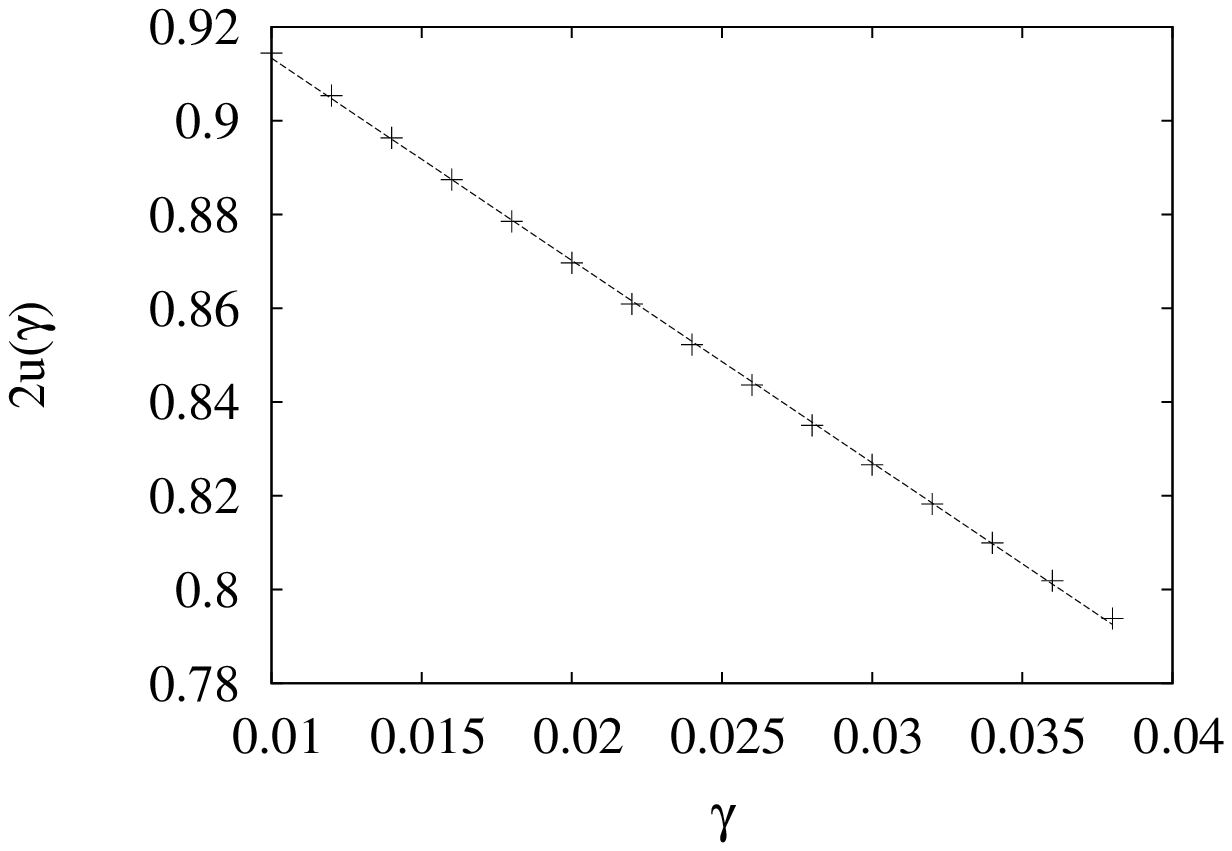}
\end{center}
\caption{Energy decay exponent. The symbols represent the results of the
fitting of the primary pulse exponential energy decay
and the line $2u(\gamma) =
0.956464 -4.31451\gamma$, which is the best linear fit of those points.}
\label{fig:b}
\end{figure}

\begin{figure}[ht]
\begin{center}
\includegraphics[width=10cm]{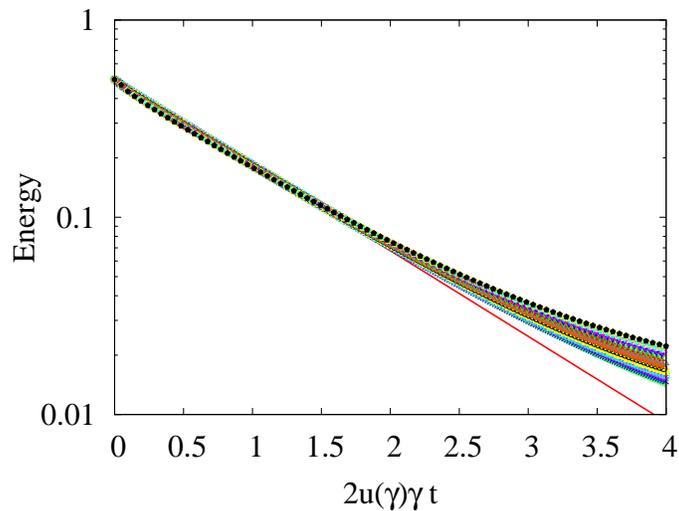}
\end{center}
\caption{(Color online) Early total energy decay for $n=5/2$. The line is the
expected behavior according to our hypothesis
$E = (1/2) \exp[-2u(\gamma)\gamma t]$, while the different symbols represent the energy decay for
different values of $\gamma$ ranging from 0.01 (lowest curve)  to 0.039 (highest curve).}
  \label{fig:energydecay}
\end{figure}

While Fig.~\ref{fig:pulseenergy} shows an exponential decay of the energy
of the primary pulse with time, 
there is a small spread in the slope of the logarithm $E_P$ vs $\gamma t$,
indicative of a mild $\gamma$ dependence of $u$ on $\gamma$. 
We thus write more accurately 
\begin{equation}
E_P \sim e^{-2u(\gamma) \gamma t}.
\end{equation}
The dependence of $u$ on $\gamma$, which is essentially
linear and indeed mild, is seen in Fig.~\ref{fig:b}.  
We also confirm our ubiquitous assumption that the total energy and the energy in the primary pulse
are essentially equal up to times of order $\gamma t\sim 1$.  This is shown
in Fig.~\ref{fig:energydecay}, where we plot the
total energy (on a logarithmic scale) vs $2u(\gamma)\gamma t$.  The deviations from this behavior
that set in beyond these times are greater for larger viscosities.  

\subsection{Secondary pulse}

Returning to Fig.~\ref{fig:snap_g002}, we follow the continuing history of the excitation.  Above
we have described the behavior of the primary pulse during the course of its existence.  The figure
illustrates the evolution of a long-lived secondary pulse during this time interval. This secondary
pulse, also being a nonlinear disturbance, continues to change in shape.  The pulse steepens (becoming
more and more asymmetric) as its peak travels faster than the 
bottom right of the peak. Note that the primary and secondary pulses have comparable amplitudes even
while they are still distinguishable before the primary pulse dissipates.  When the secondary pulse
is sufficiently steep, dispersion begins to prevail and the front displays oscillatory structure
with peaks that are a few grains wide.  This is shown in the inset of Fig.~\ref{fig:snap_g002}.  The
secondary pulse is shock-like, with velocities of the grains in the pulse at least an order of
magnitude smaller than the pulse phase speed.

\begin{figure}
\begin{center}
\includegraphics[width=9cm,height=6cm]{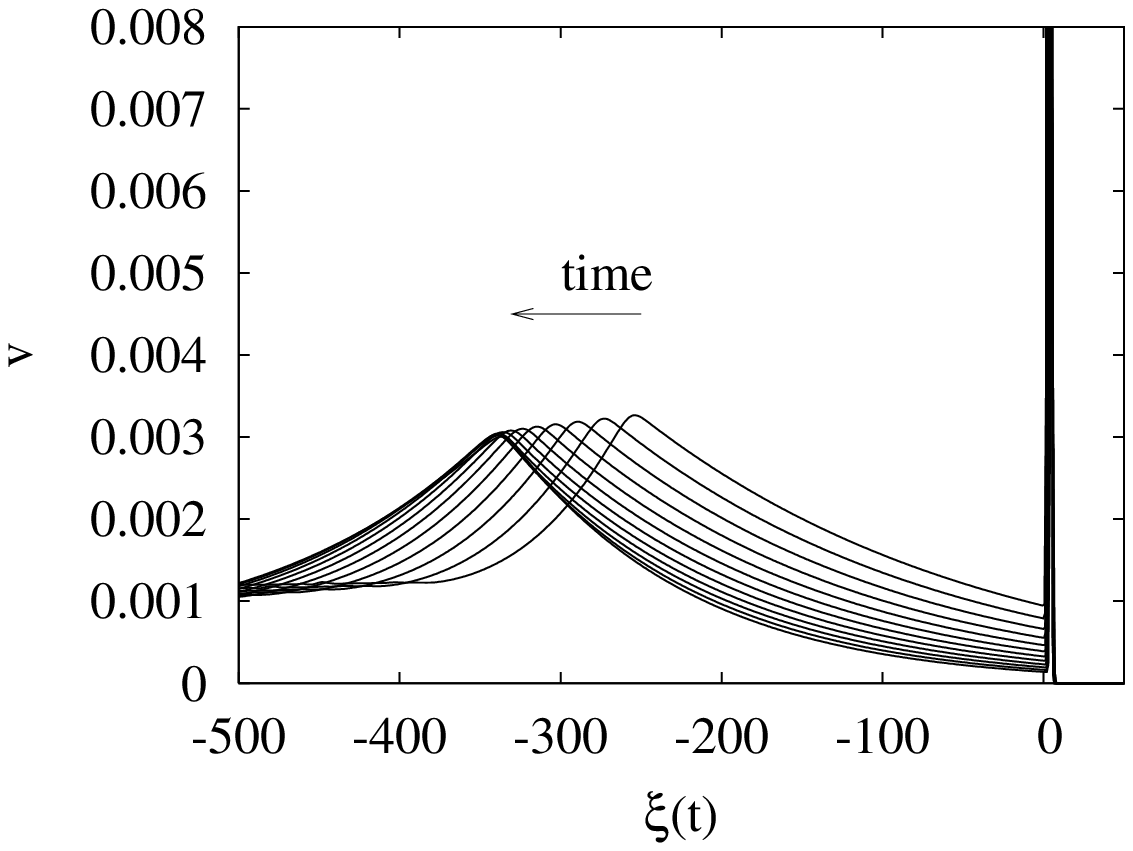}
\caption{
Snapshots of the velocities of the grains at early times ranging from
$40$ to $95$ in steps of $5$ in nondimensional units. The abscissa is the
moving variable $\xi(t)\equiv k-\int_0^t c(t)dt$, where $c(t)$ is the
time-dependent velocity of the primary pulse, and $k$ denotes the
granule in the chain.  
Eventually the primary pulse and the compression behind it vanish
and the secondary pulse
continues to move at an essentially constant velocity
($n=5/2$ and $\gamma=0.005$).}
\label{fig:secondary1}
\end{center}
\end{figure}

\begin{figure}
\begin{center}
\includegraphics[width=9cm,height=6cm]{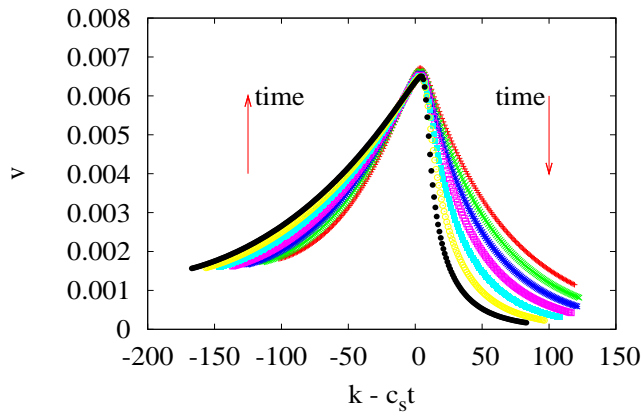}
\caption{(Color online)
Snapshots of the velocity profiles of the secondary pulse for $n=5/2$ and $\gamma=0.01$ at different
times, with increase upward on the left side of the pulse and downward on the right, consistent
with the steepening of the pulse with time.  The abscissa is now the moving variable with respect to
the essentially constant velocity of the peak of the secondary pulse.}
\label{fig:secondary2}
\end{center}
\end{figure}

Figures~\ref{fig:secondary1} and \ref{fig:secondary2} detail the behavior of the secondary pulse 
while the primary pulse has not yet disappeared.  At first the secondary pulse moves more slowly
than the primary, but this trend reverses as the primary pulse slows down with its loss of energy
and the peak of the secondary pulse acquires an essentially constant velocity.  Both figures show
that the secondary pulse is asymmetric, generating an extremely persistent tail of essentially equal
velocity granules behind it (not shown explicitly in the figures).  This asymmetry sharpens with
time, as is made evident in Fig.~\ref{fig:snap_g002}.

\begin{figure}[h]
    \begin{center}
      \includegraphics[width=10cm]{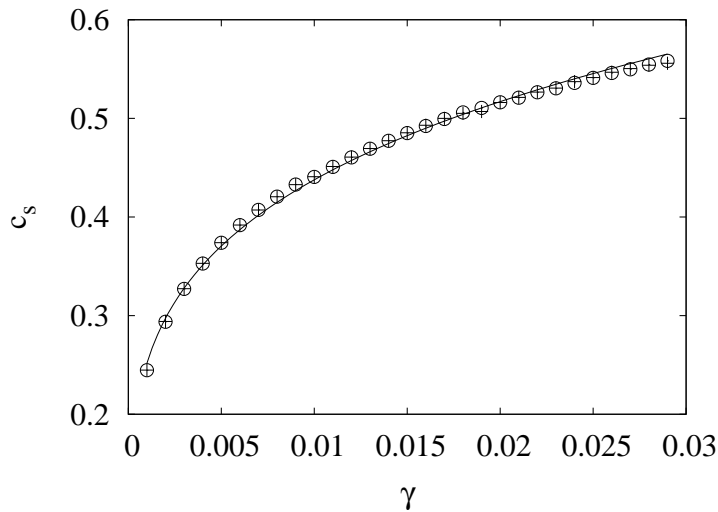}
    \end{center}
    \caption{Comparison between the sound speed given by the velocity of the maximum of the
secondary pulse (circles) and Eq.~(\ref{eq:soundspeed}) (plus signs) for $n=5/2$. The line
represents a power law best fit to the data (the difference between fitting the
circles and the plus signs is negligible). }
    \label{fig:soundspeed}
  \end{figure}

The velocity of the peak of the secondary pulse is constant throughout this early portion of its
history, as seen clearly in Fig.~\ref{fig:secondary2}, where the abscissa for all the times shown is
scaled with the same velocity $c_s$. It is furthermore extremely interesting that this peak velocity
can be associated with a ``speed of sound" in the following sense (even though this system
in the initial uncompressed state is a
sonic vacuum~\cite{nesterenko,nesterenkobook}).
That is, if $k$ denotes the position of the grain with maximum velocity,
we observe that the velocities of grains $k$ and $k+1$ are related as 
\begin{equation}
 v_k(t) = v_{k+1}(t+\frac{1}{c_s}) \approx v_{k+1}(t) +\dot{v}_{k+1}(t)
\frac{1}{c_s},
\end{equation}
where the speed $c_s$ is expressed in units of grains per unit time, and the approximation
follows from a Taylor series expansion to first order.
Therefore, the relative velocity $v^{(R)}_k = v_k - v_{k+1}$ can be written as
\begin{equation}
v^{(R)}_k = \dot{v}_{k+1} \frac{1}{c_s}.
\end{equation}
Further, we can thus write
\begin{equation}
c_s \left( v^{(R)}_k - v^{(R)}_{k-1} \right) = \dot{v}_{k+1} - \dot{v}_k.
\end{equation}
Since the medium is precompressed (because of the first pulse), we may
linearize Eq.~(\ref{eq:motion})~\cite{nesterenkobook}
\begin{eqnarray}
\dot{v}_k &=& -\Delta^{n-1} + (\Delta + v^{(R)}_{k-1} \frac{1}{c_s})^{n-1}
\simeq \Delta^{n-2}(n-1)\frac{v^{(R)}_{k-1}}{c_s},\\
\dot{v}_{k+1} &=& - (\Delta - v^{(R)}_{k} \frac{1}{c_s})^{n-1} + \Delta^{n-1}
\simeq \Delta^{n-2}(n-1)\frac{v^{(R)}_{k}}{c_s},
\end{eqnarray}
where $\Delta\equiv x_{k}-x_{k+1}$ is the compression.  Hence,
\begin{equation}
c_s \left( v^{(R)}_k - v^{(R)}_{k-1} \right) = (n-1) \frac{\Delta^{n-2}}{c_s}
\left( v^{(R)}_k - v^{(R)}_{k-1} \right),
\end{equation}
so that the sound speed is finally given by (see Eq.~(1.114) in~\cite{nesterenkobook})
\begin{equation}
  c_s = \sqrt{(n-1) \Delta^{n-2}}.
\label{eq:soundspeed}
\end{equation}
Comparison between this result and the speed of the secondary pulse maximum obtained from our
simulations during the time interval in which the primary pulse has not yet dissipated is shown in
Fig.~\ref{fig:soundspeed} for $n=5/2$. The agreement is clearly excellent, and extends to several
significant figures.  The sound speed in this case increases as $\gamma^{0.24}$.
The dramatic agreement observed in the figure is also found for other values of $n$. 
As $\gamma$ increases, the time it takes for the second pulse to ``catch" the first one 
decreases and it becomes difficult to calculate the sound speed.
Therefore, we are only able to calculate this speed for small viscosities.

\begin{figure}[ht]
  \begin{center}
    \includegraphics[width=10cm]{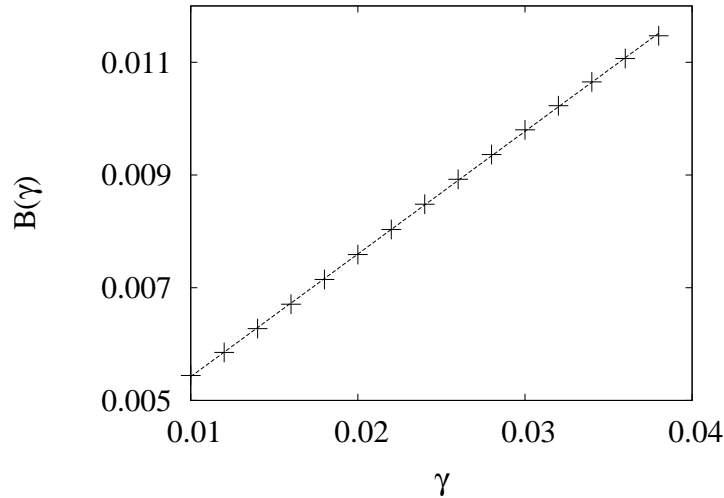}
  \end{center}
  \caption{Secondary pulse energy saturation constant for $n=5/2$. The symbols represent the
results of fitting Eq.~(\ref{eq:secpulse}), and the line  $B(\gamma) =
0.00324636 + 0.217601\gamma$ is the best linear fit for these points.}
  \label{fig:c}
\end{figure}

\begin{figure}[ht]
  \begin{center}
    \includegraphics[width=10cm]{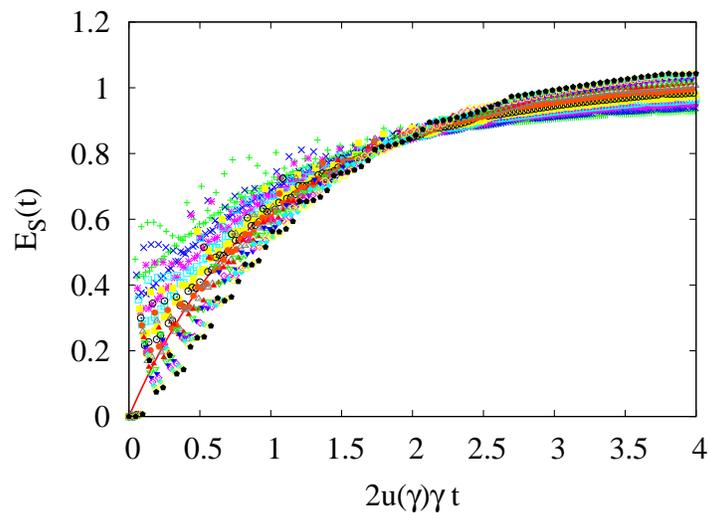}
  \end{center}
  \caption{(Color online) Collapse of the energy behind the primary pulse for $n=5/2$ and different
values of $\gamma$. The line is $1 - e^{-2u(\gamma)\gamma t}$, the behavior according
to the hypothesis Eq.~(\ref{eq:secpulse}), while the symbols represent the simulation results.}
  \label{fig:seccol}
\end{figure}

Consider next the energy in the secondary pulse.  The increase in this energy is observed to be of
the form $B\left(1-e^{-2u\gamma t}\right)$, where $B$ depends on
$n$ and $\gamma$ and is the maximum energy of the secondary pulse.
More accurately, 
\begin{equation}
  E_S(t) = B(\gamma) \left(1 - e^{- 2u(\gamma) \gamma t}\right).
  \label{eq:secpulse}
\end{equation}
In Fig.~\ref{fig:c} we plot $B(\gamma)$ as a function of $\gamma$ for $n=5/2$, showing a
fairly linear dependence. However, although the trend is captured when we plot 
$E_S(t)/B(\gamma)$ vs  $2u(\gamma) \gamma t$ as in Fig.~\ref{fig:seccol}, we do not observe a clean
collapse as predicted by
Eq.~(\ref{eq:secpulse}), indicative of a more complex $\gamma$ dependence.
In any case, the secondary pulse reaches its maximum value over a time scale of order $\gamma t \sim
1$ and then remains essentially constant over a much longer time scale, although eventually its
energy will also be dissipated.

\subsection{Higher viscosities}

The detailed results presented to this point are associated with small
values of $\gamma$.  In this regime it has been reasonable to speak of
two pulses as though they were separate entities, 
the primary being mainly due to nonlinearity and
discreteness, and the secondary caused mainly by dissipation and
nonlinearity.  Nevertheless, it is not inappropriate for these low viscosities to speak
of a ``separation" of pulses.  The primary pulse causes the
precompression that underlies the secondary pulse, and in this sense
both together are a single entity.  Summarizing this regime,
for \emph{small viscosities} ($\gamma \le 0.03$) tha
secondary pulse reaches a critical slope for transition to an
oscillatory profile \emph{before} catching the
primary pulse, while the primary pulse loses almost all of its
energy before being absorbed by the secondary pulse
(Fig.~\ref{fig:snap_g002}; the oscillatory shock profile first emerges when
the secondary pulse is in the vicinity of particle 300, not shown
in the figure).  Note that a critical viscosity for a transition from an oscillatory
to a monotonous profile of a shock wave is found in~\cite{herbold2}.
We have observed that in this small-$\gamma$ regime 
the maximum velocity in the secondary pulse increases
with increasing viscosity because larger dissipation is associated with a
greater compression, resulting in a secondary pulse of higher
amplitude.  For very small
$\gamma$ ($\lesssim 0.002$) the secondary pulse has an almost
imperceptible amplitude on our numerical scale (and of course it
disappears entirely when $\gamma=0$), and the primary pulse has a
very long life. However we do not find a transition to a regime without
a secondary pulse for any finite value of $\gamma$.
The secondary pulse fades away smoothly with diminishing $\gamma$.

\begin{figure}
\begin{center}
\includegraphics[width=8cm,height=6.5cm]{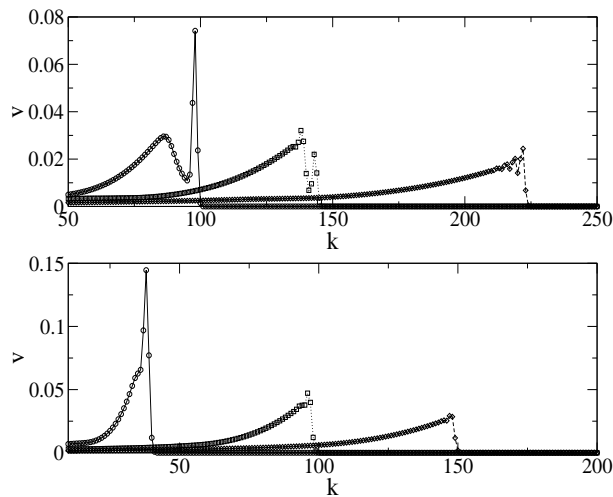}
\caption{
Upper panel:
snapshots of the velocity profile for intermediate viscosity
($\gamma=0.04$, $n=5/2$) at
different times: $140$, $220$,  and $400$.
Lower panel: snapshots of the velocity profile for large viscosity
($\gamma=0.1$, $n=5/2$) at different times: $100$, $300$, and $500$. 
\label{fig:snap_g004_g01}}
\end{center}
\end{figure}

For \emph{intermediate viscosities} ($0.04 \le \gamma \le 0.07$) the
secondary pulse catches up with the primary pulse while the primary
pulse still has an amplitude comparable to the secondary
(upper panel of Fig.~\ref{fig:snap_g004_g01}). As in the previous case,
after the first pulse
disappears, the secondary pulse propagates as a shock-like wave with
an oscillatory front caused by the dispersion.

\begin{figure}[h]
    \begin{center}
\includegraphics[width=10cm]{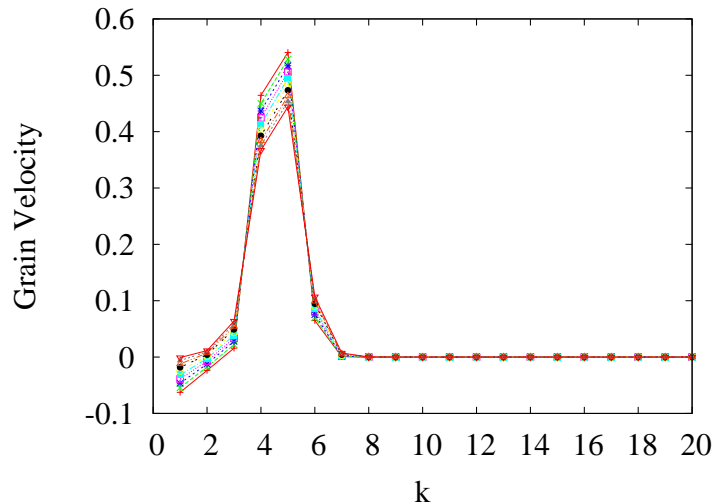}
    \end{center}
    \caption{(Color online) Early dynamics. The first pulse always occurs at early times.
At the beginning, the only role of the viscosity is to dissipate energy. The
shape of the pulse and the time it takes to develop it is the same for $\gamma$
ranging from 0.01 to 0.1}
\label{figure13}
  \end{figure}

For \emph{large viscosities} ($\gamma \ge 0.07$) there
is no clear distinction between the primary and secondary pulses.
Actually, for viscosities $\gamma \ge 0.1$ it is no longer
appropriate to think of two separate pulses
(lower panel of Fig.~\ref{fig:snap_g004_g01}).  From \emph{almost} the beginning, there
is a single shock-like structure of dissipative origin with a sharp
monotonic front.  Nevertheless, it should be noted that the first pulse is \emph{always} evident,
albeit for a very short time.  In Fig.~\ref{figure13} we see that it takes about four granules for it to
develop fully, and that the dynamics up to 10 grains or so is
always the same, that is, the shape of the pulse and the time it
takes to develop are essentially the same for viscosities ranging from 0 to 0.1. After this early
time either the second pulse appears behind the first pulse (small $\gamma$), or the first pulse
appears to be essentially deformed into the second pulse (larger $\gamma$).  

\section{Dynamics with precompression and with initial gaps}
\label{sec3}
The results presented up to this point deal with a chain in which the granules are initially placed
side by side, with no precompression and with no gaps (i.e., a sonic
vacuum~\cite{nesterenko,nesterenkobook}).  In this section 
we explore the consequences of relaxing these assumptions.  In particular, 
we explore the way in which a variation in the initial placement of granules modifies or
otherwise distorts the two-component wave disturbance discussed above.  Based on numerical
simulations, we present these effects through a series of figures.

\begin{figure}[h]
    \begin{center}
\vspace*{1.5in}
\includegraphics[width=10cm]{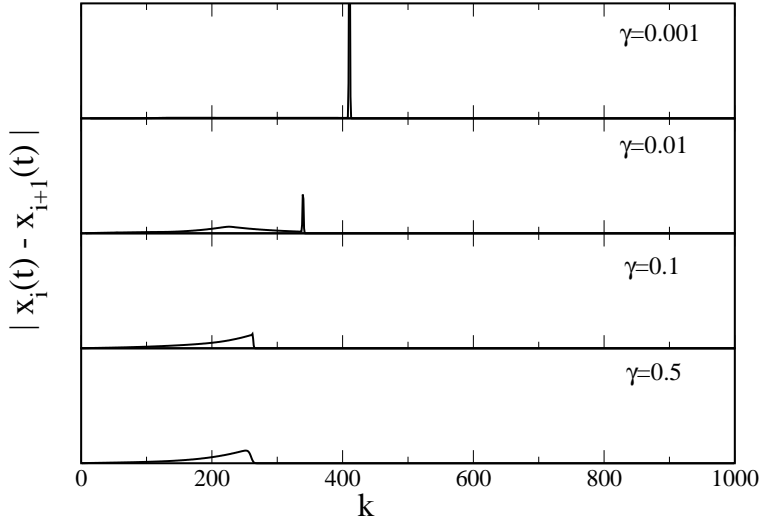}
    \end{center}
    \caption{Grain compression along the chain with no precompression at 50,000 iterations. Panel
progression shows the chain at the same instant of time under the effect of increasing dissipation.
}
\label{figure14}
  \end{figure}

The first of these figures, Fig.~\ref{figure14}, is simply a recap of previous results that facilitates our
discussion.  In this figure there is no precompression, and we show the grain compression 
wave structure at a particular instant of time after 50,000 iterations as the
damping increases.  At extremely low damping there is a solitary wave accompanied by an extremely
low amplitude secondary wave that is not visible on the scale of the figure.  As the viscosity increases,
the secondary pulse behind the primary pulse becomes more prominent.  Increasing the viscosity
further leads to a single shock-type pulse (the primary pulse has already disappeared), with a wider
shock front at the larger viscosity.

\begin{figure}[h]
    \begin{center}
\includegraphics[width=10cm]{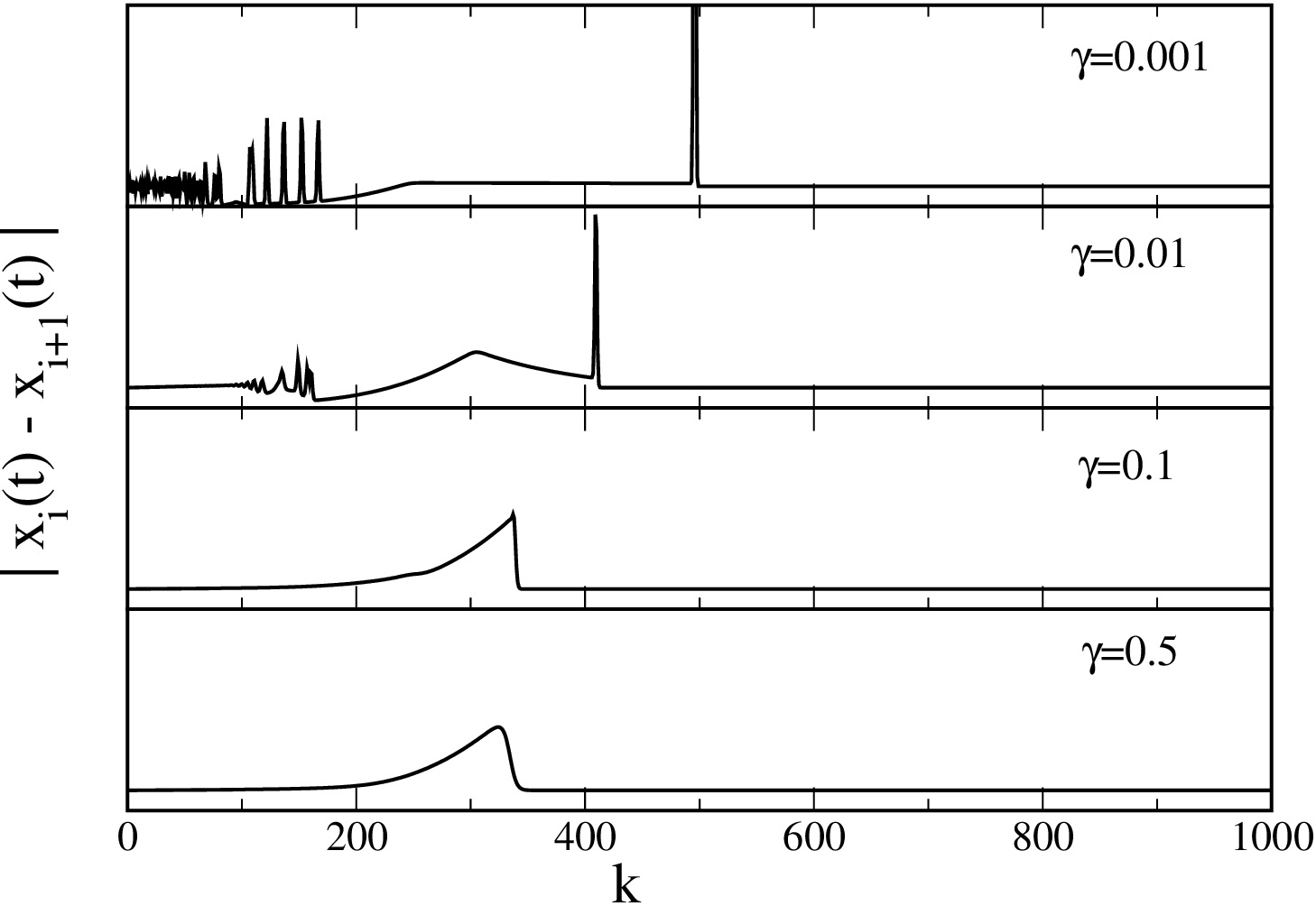}
    \end{center}
    \caption{Pulse propagation in a precompressed chain in which the precompression is 100 times
weaker than the dynamic displacement imparted by the initial pulse. The scale on the $y$ axis in
each panel goes from $0.0$ to $0.1$, and viscosity increases downward, as indicated.  
}
\label{figure15}
\end{figure}

\begin{figure}[h]
    \begin{center}
\vspace*{0.5in}
\includegraphics[width=10cm]{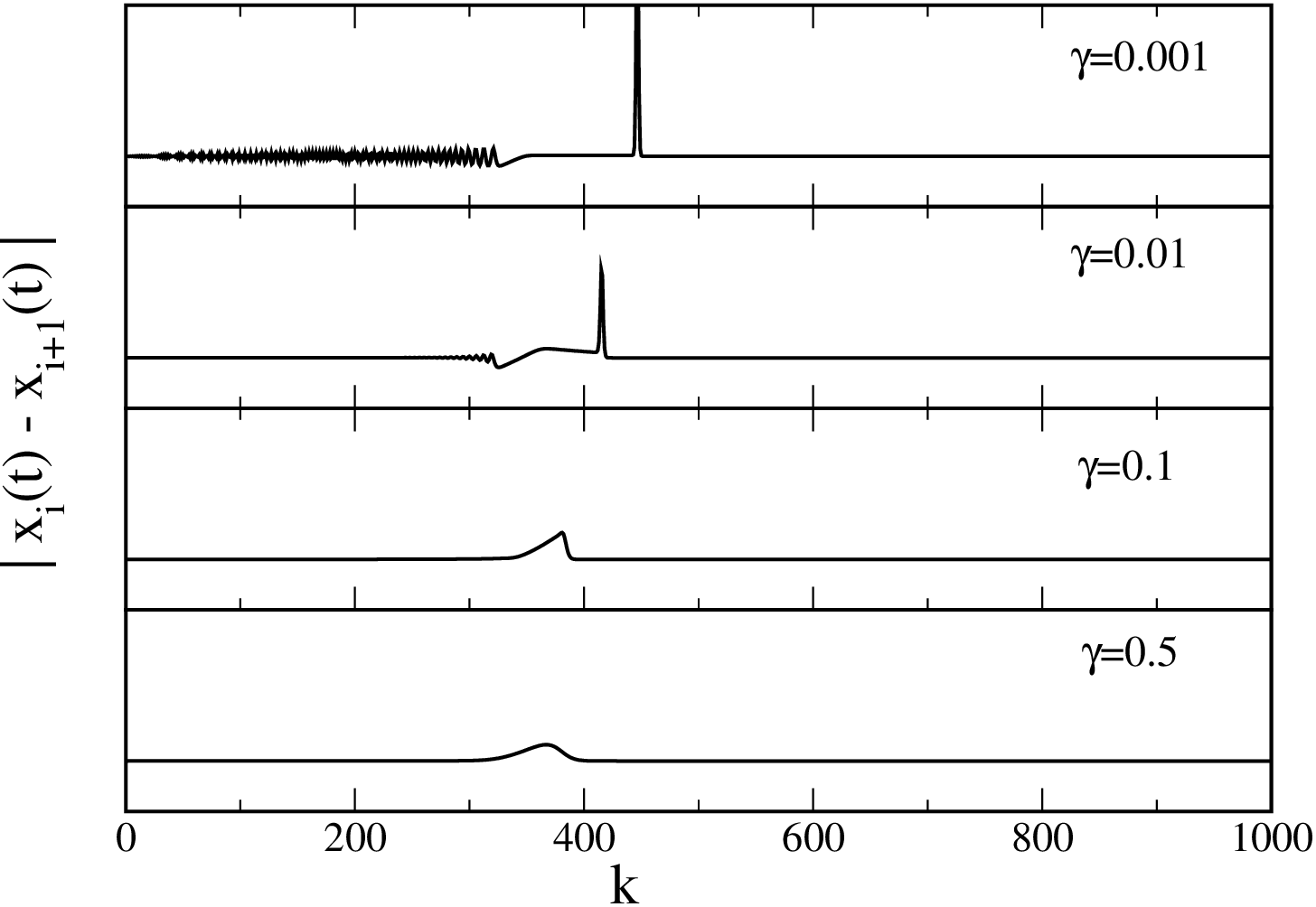}
    \end{center}
    \caption{Pulse propagation in a precompressed chain in which the precompression is 10 times
weaker than the dynamic displacement imparted by the initial pulse. The scale on the $y$ axis in 
each panel goes from $0.0$ to $0.4$, and viscosity increases downward, as indicated.  
}
\label{figure16}
\end{figure}

\begin{figure}[h]
    \begin{center}
\includegraphics[width=10cm]{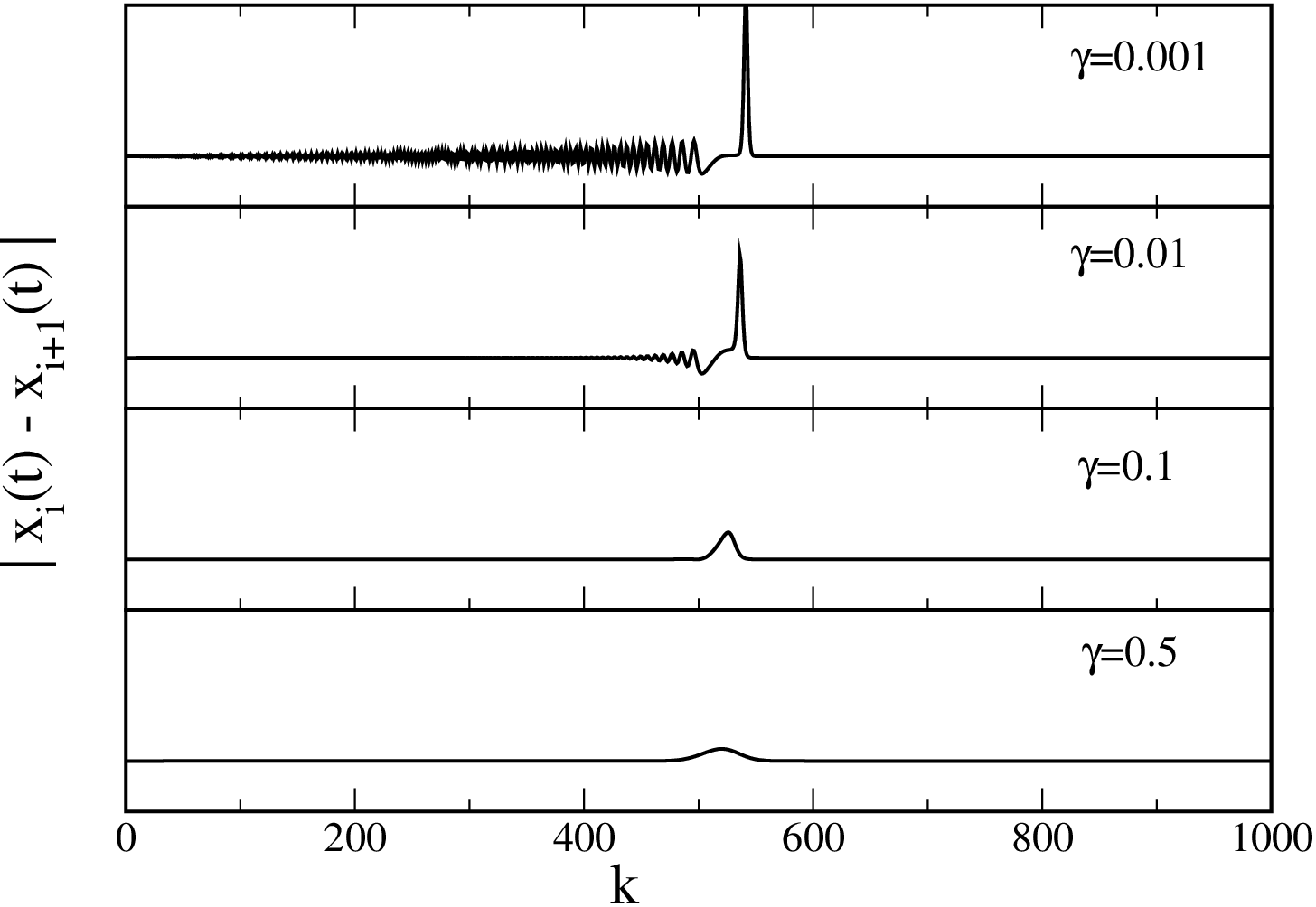}
    \end{center}
    \caption{Pulse propagation in a precompressed chain in which the precompression is half as large
as the dynamic displacement imparted by the initial pulse. The scale on the $y$ axis in
each panel goes from $0.0$ to $0.8$, and viscosity increases downward, as indicated.  
}
\label{figure17}
\end{figure}

Next we explore the consequences of precompression, which qualitatively changes our system from a
sonic vacuum to a more traditional discrete system with finite sound speed. We ensure mechanical equilibrium
by applying appropriate static forces to the end particles of the chain.
The next three figures show progressions of increasing viscosity at a given precompression. 
Viscosity increases from one figure to the next (note the different $y$-axis scales in the three
figures).  Figure~\ref{figure15} shows the changes introduced by a very
small precompression.  The features illustrated in Fig.~\ref{figure14} are repeated in this figure
at the same instant of time, specifically the presence of a primary and a secondary pulse 
when the viscosity is small, and, at higher viscosities the single shock-type pulse
of increasing front width as viscosity increases.  The new feature here, namely, the rarefaction
wave and the tail of attenuating compression peaks with zero minimum amplitude that follow the
secondary wave, are due to the precompression.  They occur at zero viscosity and survive the
effects of low viscosities. The attenuating compression peaks of zero minimum amplitude are
related to the rattling of partciles at the impacted end accompanied by the opening and closing
of gaps. In Figs.~\ref{figure16} and \ref{figure17} we see that increasing precompression 
causes the secondary wave (when it is visible at all) to decrease in amplitude and width, but that
the other features (primary wave with trailing rarefaction wave and attenuating oscillatory tail at
lower viscosities, single shock type pulse at higher viscosities) continue to persist.  Note that
greater precompression leads to a stronger (and therefore faster) primary pulse, an effect that has
nothing to do with the viscosity but that we just point to for completeness of the description.

Finally, we recall that with weak or zero precompression the velocity of the primary excitation is
related to its amplitude in a nonlinear fashion determined by the nonlinearity of the medium
(and of course by the dissipation that causes a decrease in the amplitude with time) and
essentially unrelated to any speed of sound in the medium. On the other hand, the secondary pulse
velocity is related to the compression of the medium that follows the primary pulse and is 
determined by the speed of sound. With increasing precompression, the speed of sound increasingly
determines the speed of disturbance propagation, as we can see, for example, in Fig.~\ref{figure17},
where all four panels show essentially the same speed.  The larger the precompression, the smaller
the effect of the nonlinearity; at sufficiently high precompression relative to the compression
caused by the initial pulse, the medium effectively becomes linear.

Finally, the introduction of relatively large gaps in the initial distribution of granules causes the
system to no longer behave as a collective entity and it will instead resemble a granular gas.
We have determined that gaps up to size of order $10^{-5}$ in our dimensionless units lead to
behavior similar to that of the original system.  However, beyond this the loss of energy and the
character of the excitation begin to change drastically.

\section{Summary}
\label{sec4}
In this work we have explored the interplay of nonlinearity, discreteness, and dissipation in a
one-dimensional dense granular chain.  In particular, we have
expanded on our earlier demonstration of an interesting two-wave structure
observed in a dissipative granular chain excited by a $\delta$-force applied to a single
grain~\cite{ourprl}.  One might expect a one-wave structure such as an attenuating
solitary wave or an attenuating shock wave -- the observation of a complex two-wave structure is 
certainly unusual.  This structure, which occurs at low viscosities (but \emph{requires} non-zero
viscosity), consists of a primary wave characteristic of a discrete nonlinear nondissipative sonic
vacuum, and a secondary shock-like long wave due entirely to intergrain contact viscosity. 
The high velocity gradients in the narrow primary pulse leads to its relatively rapid
attenuation, which we have shown numerically to be exponential. We have furthermore shown that the
decay rate is essentially proportional to the viscosity $\gamma$, with small correction terms of
$O(\gamma^2)$.  During its lifetime, the speed of
the primary pulse is related to its amplitude in the same way as in a nondissipative
chain~\cite{nesterenko,nesterenkobook,hinch,ourpre}, but now they both decrease with time as the
primary pulse dissipates.  Some of the dissipated energy is simply lost, while some of the initial
energy of the primary pulse is
transferred to the secondary pulse caused by the compression left behind by the primary pulse (which
would not occur in the absence of viscosity).  The much smaller velocity gradients in the secondary
pulse cause it to be very long-lived, and its speed is essentially the local speed of sound. Below a
critical viscosity the secondary pulse develops a dispersion-induced oscillatory front.  There are
thus three distinct time scales in this problem: an extremely short scale for the formation of the
primary pulse, a relatively rapid time scale of attentuation of the primary pulse, and a very slow time
scale for the eventual attenuation of the secondary pulse.  At higher viscosities it becomes less
and less appropriate to think of the primary and secondary pulses as separate entities. Instead, one
observes a spatially lengthening excitation that presents a monotonic front.  This description is
appropriate after a very short time during which there is always a primary pulse, with
formation and evolution characteristics essentially independent of the viscosity.  But this primary
pulse is dissipated very quickly at higher viscosities and most of the dynamical regime is dominated
by the shock-like wave.  We have provided a particularly detailed picture of
the formation and evolution of the structure when the grains are initially in contact with one
another but without precompression (sonic vacuum). 

We also explored the consequences of relaxing the initial configuration, particularly to the case of
initial precompression.  We found that as long as the precompression is small compared to the
dynamical compression produced by the initial velocity impulse, the features described above are still
observed, albeit somewhat complicated by the appearance of a rarefication wave and a tail of
attentuating compression peaks of zero minimum amplitude behind the secondary wave.  As before,
there are distinct primary and secondary waves when the viscosity is low, and (except for very short
times) a single shock-type pulse when the viscosity is high.  We also observed that increasing
precompression leads to increasingly linear behavior in which any oscillatory features decrease in
amplitude, and the excitation propagates at essentially the speed of sound in the medium. 

Finally, we briefly explored the opposite relaxation of the initial condition, namely, the effect of
initial gaps between the grains.  When these gaps are extremely small, of size $10^{-5}$ or smaller
in our dimensionless units, the behavior is similar to that of the chain without gaps, but beyond
this the character of the dynamics of the excitation changes drastically.  We have not explored
this regime in detail.

We expect the two-wave phenomenon involving vastly different length and time scales to occur in
other nonlinear discrete dissipative systems under conditions of short pulse loading.  Examples
might include femtosecond-laser generated pulses, waves generted in atomic lattices by bombardment
of low density beams of ions, and waves in three-dimensional packings of spherical beads immersed in
liquid under short-duration plane explosive loading.  While our analysis has focused on
one-dimensional chains, we expect the phenomenon to occur in systems of higher dimensions as well,
where issues of geometry of the constituents and of the initial loading introduce additional
interesting variables.  Work along these various directions is in progress.

\section*{Acknowledgments}
Acknowledgment is made to the Donors of the American Chemical Society Petroleum Research Fund for
partial support of this research (K.L.).
This work was supported in part by the Conselho Nacional de Desenvolvimiento Cient\'{\i}fico e
Tecnol\'ogico (CNPq) (A.R.), CONACyT Mexico Projects No. J-59853-F and 48783-F (A.H.R.), and NSF Grant No.
DCMS03013220 (V.F.N.).

\end{document}